\documentclass[12pt,preprint]{aastex}

\shorttitle{On the Age of the Widest VLM Binary}
\shortauthors{Artigau et al.}

\begin{document}

\title{On the Age of the Widest Very Low Mass Binary}

\author{\'Etienne Artigau\altaffilmark{1}, David Lafreni\`ere\altaffilmark{2}, Lo\"{\i}c Albert\altaffilmark{3} and Ren\'e Doyon\altaffilmark{4}}

\altaffiltext{1}{Gemini Observatory, Southern Operations Center, Association of Universities for Research in Astronomy, Inc., Casilla 603, La Serena, Chile}
\altaffiltext{2}{Department of Astronomy and Astrophysics, University of Toronto, 50 St. George Street, Toronto, ON M5S 3H4, Canada}
\altaffiltext{3}{Canada-France-Hawaii Telescope Corporation, 65-1238 Mamalahoa Highway, Kamuela, HI 96743}
\altaffiltext{4}{D\'epartement de Physique and Observatoire du Mont M\'egantic, Universit\'e de Montr\'eal, C.P. 6128, Succ. Centre-Ville, Montr\'eal, QC, H3C 3J7, Canada}

\email{eartigau@gemini.edu, lafreniere@astro.utoronto.ca, albert@cfht.hawaii.edu, doyon@astro.umontreal.ca}

\begin{abstract}
We have recently identified the widest very low-mass binary 
(2M0126AB), consisting of an M6.5V and an M8V dwarf with a
separation of $\sim$5100~AU, which is twice as large as that of the
second widest known system and an order of magnitude larger than those
of all other previously known wide very low-mass binaries. If this
binary belongs to the field population, its constituents would have
masses of $\sim$0.09 M$_{\odot}$, at the lower end of the stellar
regime. However, in the discovery paper we pointed out that its proper
motion and position in the sky are both consistent with being a member
of the young (30~Myr) Tucana/Horologium association, raising the
possibility that the binary is a pair of $\sim$0.02~M$_{\odot}$ brown
dwarfs. We obtained optical spectroscopy at the Gemini South Observatory in
order to constrain the age of the pair and clarify its nature. The
absence of lithium absorption at 671~nm, modest H$\alpha$ emission, and the
strength of the gravity-sensitive Na doublet at 818~nm all point
toward an age of at least 200~Myr, ruling out the possibility that the binary is a member of Tucana/Horologium. We further estimate that the binary
is younger than 2~Gyr based on its expected lifetime in the galactic
disk.

\end{abstract}
\noindent{\em Suggested running page header:} 

\section{Introduction}
Binarity is ubiquitous in stellar systems, from the most massive stars
to the substellar regime. The statistical properties of binaries 
and higher multiple systems retain information on the physical
processes that led to their formation long after they have 
dispersed beyond their star-forming regions. While nearly all early-B
stars reside in binary systems, with a binarity fraction as high as
$\sim$$80\%$ \citep{Kouwenhoven2005}, this fraction falls to as low as
$\sim$$20\%$ \citep{Burgasser2007} for brown dwarfs. Also, while stellar binaries are found with separations reaching 35~000~AU for 
early-B primaries and 20~000~AU for A primaries \citep{Abt1988}, only a handful of substellar systems have
separations beyond 200~AU. Most of these have been found in
star-forming regions and open clusters, such as Oph 1623-2402 (212~AU)
and Oph 1622-2405 (243~AU) in the star-forming clouds of Ophiuchus
\citep{Close2007}, 2M1101-7732 (241.9~AU) in the Chamaeleon I star-forming
region \citep{Luhman2004} and the system SE70/S Ori68, a likely
$1700\pm300$~AU-wide system in the $\sigma$~Orionis cluster
\citep{Caballero2006}. Only three wide very low mass (VLM) binaries
are known in the field, DENIS055-44 \citep{Billeres2005}
($\sim$$220$~AU), K\"onigstuhl 1 AB ($1800\pm170$~AU)
\citep{Caballero2007a} and 2M0126AB ($5100\pm400$~AU)
\citep{Artigau2007}. \citet{Caballero2007b} established that wide
late-type binaries with a mass ratio \textgreater$0.5$ are rare objects
in the field, with only $1.2\pm0.9\%$ of VLM stars and BDs residing in
such systems.

The 2M0126AB system represents the most extreme case of VLM binary
known. It was discovered as a common proper motion pair by \citet{Artigau2007}, 
and later verified by \citet{Caballero2007b}. The derived probability of having, 
over the whole sky, a single pair of unrelated late-Ms with matching proper motion and distances within
our uncertainties is $\sim$$0.002$ \citep{Artigau2007}; hence the two 
components are almost certainly gravitationally bound. Its components have spectral types of M6.5
and M8 as confirmed by GNIRS \citep{Elias2006} spectroscopy. While
near-infrared spectroscopy shows that K\textsc{I} equivalent widths
are compatible with those of field objects, the pair falls near the
core of the Tucana/Horologium (Tuc/Hor) \citep{Zuckerman2004} association and
shares its bulk motion, suggesting that 2M0126AB may be a member of
this $30$~Myr old moving group. If indeed it is a member of this
group, the low temperatures of its components combined with a very
young age would imply  masses at the lower limit of the brown dwarf
realm. Otherwise, if this pair is a field object (i.e. an age greater
than 1~Gyr) the models of \citet{Chabrier2000} indicate masses of
0.095~M$_{\odot}$ and 0.092~M$_{\odot}$, close to the lower limit of
the stellar regime.

Here we present new optical spectroscopy of both components of
2M0126AB to constrain the properties of this odd pair, either as a
member of a young association or as a much older field object. The
wavelength interval selected contains important age or gravity
indicators: the lithium feature at 670.9~nm, H$\alpha$ and the 820~nm
Na doublet. Observations and data reduction are described in
\S~\ref{Observations}. Results and a discussion on the physical
properties constrained by these observations are given in
\S~\ref{discussion}.

\section{Observations and data reduction} \label{Observations}

The dataset described here was obtained with the GMOS-S spectrograph
at the Gemini South telescope on 2007 October 3. A $0.5\arcsec$-wide
slit was used with the R400 grating and OG515 filter; the resulting
resolving power was R$\sim$1900 for a $600 - 950$~nm wavelength
coverage. The slit was aligned as to obtain a spectrum of both objects
simultaneously and the observations were made using the
nod-and-shuffle mode for improved sky-line subtraction. Two
44.7-minute and one 12.2-minute exposures were obtained for a total 
integration time of 101 minutes. The observations were obtained at 
a mean airmass of 1.5 under photometric conditions. The exposures were taken at
slightly different grating angles providing a spectral dithering in
order to fill the $3$~nm wavelength-coverage gaps caused by the spacing 
between the three GMOS detectors. The
790~nm, 800~nm and 810~nm central wavelengths for each exposure
displaced these gap in the $716-740$~nm and $870-882$~nm intervals,
i.e. in regions devoided of diagnostic spectral features. A white
dwarf (EG131) was observed on 2007 September 11 with the same setup
for instrumental and telluric corrections. 

The individual nod-and-shuffle frames were first dark and bias
subtracted. Each nod-and-shuffle frame was then calibrated using flats taken immediately before (790~nm setup) or after
(800~nm and 810~nm setup) the science observation. The two
nod-and-shuffle frames were then pair-subtracted, and corrected for
trace distortion. Wavelength calibration was performed using
copper-argon lamp spectra taken as part of the night-time
calibrations. Spectra were extracted from both the positive and
negative nod-and-shuffle traces for each component, resulting in a
total of 6 spectra that were interpolated on a finer common-wavelength
grid and median combined. The final signal-to-noise ratio per resolution 
element ranges from about 15 around 650~nm to 70 around 800~nm for 2M0126A.
The signal-to-noise obtained for 2M0126B is $\sim$20$\%$ lower.

An image of the 2M0126AB field was taken prior to the spectroscopic
observation. This 30.5-s $i$-band image used the GMOS-S central CCD
without pixel binning, providing a $2.5\arcmin\times5.6\arcmin$
FOV with a $0.073\arcsec$ pixel sampling. The image was bias
subtracted, fringe corrected and flat-fielded. This image was used to
search for additional comoving objects and test the possibility that one or both
components of the system is a tight binary itself.

\section{Results \& Discussion} \label{discussion}
\begin{figure}[!htbp]
\plotone{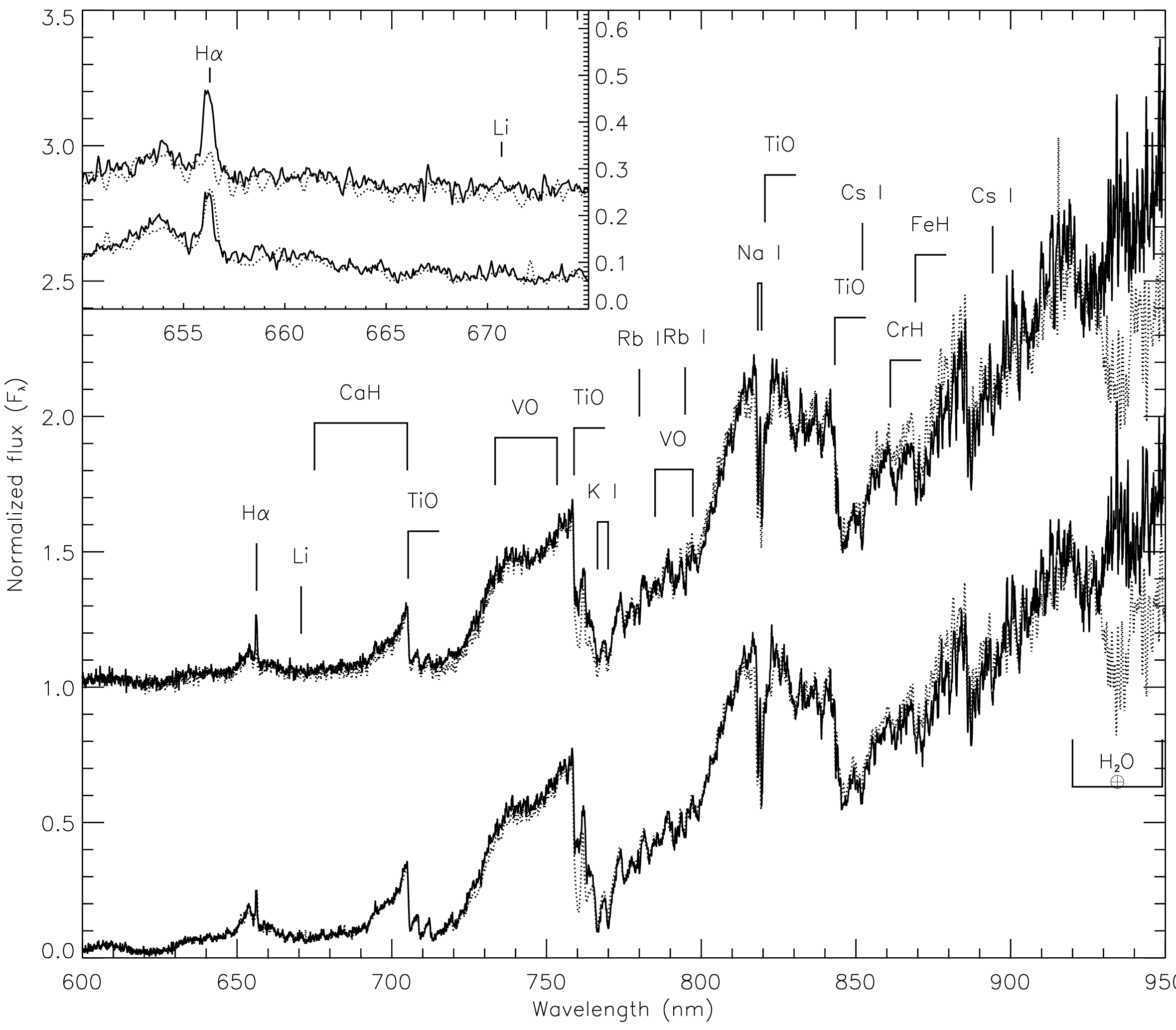}
\epsscale{.7}
\caption{\label{fig1} Spectra of 2M0126A (bottom) and 2M0126B (top,
  offset by +1) as well as two template objects, a M6.5 (2MASS
  J02422+1343, dotted line, overplotted on 2M0126A) and an M7.5 (2MASS
  J2585+1520, dotted line, offset by +1, overplotted on 2M0126B). All
  spectra have been normalized to their median flux over the
  810-840~nm spectral interval. The inset shows the 650-675~nm
  interval that contains two important age diagnostic features, the
  undetected lithium line at 670.8~nm and H$\alpha$ at
  $656.3$~nm. The inset normalization is the same as that of the main
  figure but the spectra have been offset by 0.2 for clarity.}
\end{figure}

Figure~\ref{fig1} shows the spectra of 2M0126A and 2M0126B. Both
spectra show the hallmark absorption features of late-M dwarfs such as
TiO, K\textsc{I} and Na\textsc{I}. A comparison with template M6.5 and
M7.5 spectra from \citet{Kirkpatrick1999} shows little difference with typical field objects, the
differences around 760~nm and between 930~nm and 950~nm being due to
telluric absorption (corrected in our spectra, unaccounted for in the
reference spectra).

We derive an optical spectral type using the \citet{Martin1999}
polynomial relations for the PC3 index for both components. These
spectral types agree within 0.2 subclass to those measured in \citet{Artigau2007}
and therefore we keep spectral types of \hbox{M$6.5\pm0.5$} and \hbox{M$8.0\pm0.5$}
for our analysis. Both 2M0126A and 2M0126B show moderate H$\alpha$ emission (see inset
in Figure~\ref{fig1}) with pseudo-equivalent widths (pEW) of
$-3.44$~\AA~and $-7.32$~\AA, respectively. This measurement was
converted into $\log \left( L_{\rm{H}\alpha}/L_{\rm{bol}} \right)$ using the
procedure described by \citet{Walkowicz2004}; we obtained
$\log \left( L_{\rm{H}\alpha}/L_{\rm{bol}} \right) = -4.5$ and $-4.4$ for components
A and B, respectively. The measured H$\alpha$ pEW for both components
are typical for objects of these spectral types \citep{West2004}. From
these values alone we cannot rule out the possibility that the pair is
a member of the Tuc/Hor association as young late-Ms have widely
varying levels of activity; the level observed for 2M0126A is
comparable to the least active members of the $50$~Myr old open
cluster IC~2391 while 2M0126B has H$\alpha$ strength similar to that of 
other late-Ms in the sample of \citet{Barrado2004}. Measurements of
the H$\alpha$ pEW therefore provides little help in establishing the
true nature of 2M0126AB, excluding neither of the hypotheses.

The absence of lithium absorption gives the best clue that 2M0126AB is
not a member of the Tuc/Hor association (see inset in
Figure~\ref{fig1}). At a bolometric luminosity of
$\log \left( \textrm{L}/\textrm{L}_\odot \right) \sim -3.1$, both 
objects should take
200~Myr \citep{Dantona1997,Chabrier2000} to destroy lithium, 
implying a minimum age
that is nearly an order of magnitude greater than the 30~Myr of Tuc/Hor. 
\citet{Kirkpatrick2006, Kirkpatrick2008} mention that in some young
objects, the lithium signature can be absent due to very low gravity
and the intrinsically weaker alkali lines seen in these objects. This
would explain the seemingly contradictory spectroscopic features of
2MASS J01415823-4633574: no detectable lithium at 671~nm despite
numerous low gravity indicators pointing toward an age of
5-10~Myr and an estimate gravity of $\log \left( g \right) \sim
4.0\pm0.5$. \citet{Chabrier2000} models predict, for objects at the luminosities of
2M0126A and 2M0126B, $\log \left( g \right) =4.2$ at an age of 
30~Myr (Tuc/Hor scenario), $\log g = 5.0$ at 200~Myr (minimum age with
lithium depletion) and $\log \left( g \right) =5.3$ at 1~Gyr 
(half-life of the binary due to tidal disruption in the galactic disk). While 
the $\log \left( g \right)$ predicted for 2M0126A and B in the 
Tuc/Hor scenario is close to that of 2MASS J01415823-4633574, which may cast a doubt on the
validity of the lithium test in this case, we note that objects later
than M5 in the $50$~Myr-old open cluster IC2391 do show significant
lithium absorption \citep{Barrado2004} and that, at the same
temperature as 2M0126A and B, IC~2391 late-Ms have surface gravities
within about 0.4~dex of that predicted for 2M0126A and 2M0126B at
30~Myr.

The Na \textsc{I} doublet at 818~nm is a good gravity indicator
\citep{Kirkpatrick1991, Kirkpatrick2006}. The NaI doublet equivalent
widths of \hbox{$7.04\pm0.12$~\AA} and \hbox{$6.69\pm0.20$~\AA} for 2M0126A and B
respectively, are consistent with those of field objects but about
2~\AA~larger than the values found for late-M Pleiades members
\citep{Martin1996}. This indicates that both objects are significantly
older than the Pleiades (130 Myr; \citet{Barrado2004}), which in turn implies
surface gravities of $\log \left( g \right) > 5.0$. Such a high surface gravity further
implies that the binary system would clearly show Li absorption should it be
present. This confirms the lower limit on the age set by the lithium
test at 200~Myr. We therefore establish that 2M0126AB cannot be a
member of the Tuc/Hor association and is more likely a field
pair. 

Interestingly, this pair has an expected survival time in the
galactic disk significantly shorter than the age of the disk
itself. This timescale has been estimated by scaling to 2M0126AB the
\citet{Weinberg1987} results for binaries in the solar
neighborhood. With an half-life of $\sim$1~Gyr, we can estimate at
the 68\% confidence level (equivalent to 1$\sigma$) that the pair has
an age between 0.2 and 1.8 Gyr. Masses have been determined from the \citet{Chabrier2000} models for ages of 0.2~Gyr
and $>$1~Gyr and are given in Table~\ref{tbl-1}.

\begin{figure}[!ht]
\plotone{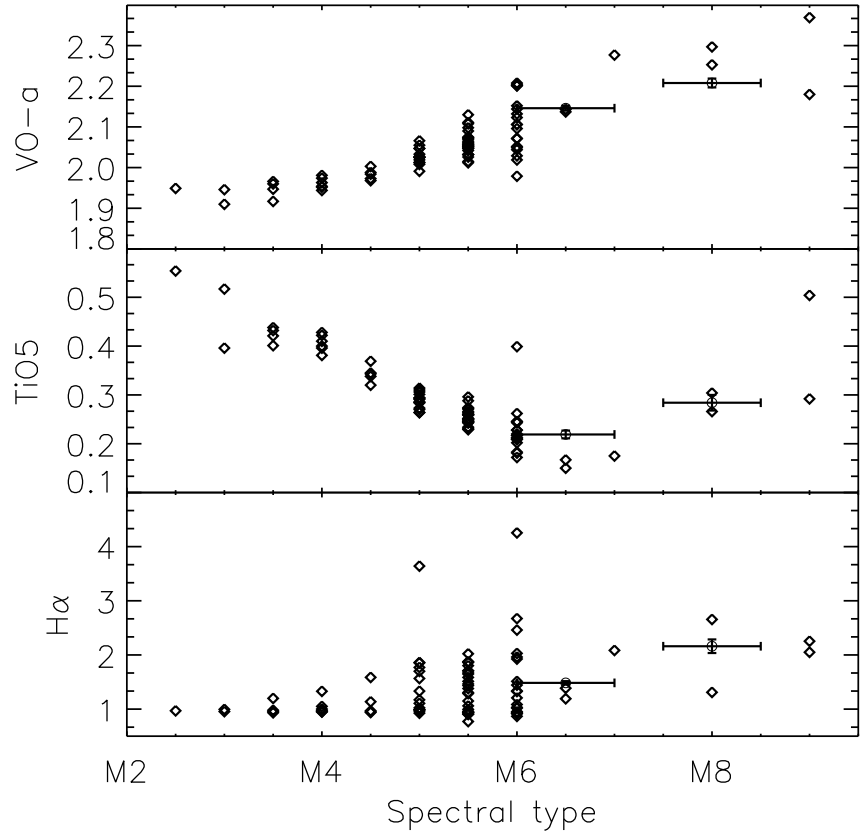}
\epsscale{.7}
\caption{\label{fig2} VO-a, TiO5 and H$\alpha$ indices as a function
  of spectral type for the \cite{Cruz2002} sample compared to the
  values measured for 2M0126A and 2M0126B (data points with error
  bars). The values measured for our binary are consistent with the
  overall trends of the field M dwarf sample, indicating neither
  particularly large chromospheric activity nor signs of peculiar
  metallicity.}
\end{figure}

We measured a list of 12 spectroscopic indices (see table~\ref{tbl-1})
compiled by \citet{Cruz2002} in order to verify whether 2M0126AB shows
any peculiarity compared to field M dwarfs. All indicators fall within
the trends observed for field objects; three are shown in
Figure~\ref{fig2}. None of the indicators deviate from the overall
trends for field M dwarfs. In particular, the TiO5, being much
stronger in sub-dwarfs and extreme sub-dwarfs, is a useful probe of
metallicity. Our CaH1, CaH2 and CaH3 versus TiO5 measurements clearly
put 2M0126A in the non-subdwarf part of Figure~1 in \citet{Gizis1997},
which extends only up to a spectral type of M7. This is in agreement
with the $\lesssim$$2$~Gyr age expected from dynamical
considerations. Using the relations between the TiO5 and CaH2 indices
versus the $J$-band absolute magnitude given by \citet{Cruz2002}, we
derive a photometric distance of \hbox{$82\pm5$~pc} for 2M0126A. The
photometric distance for 2M0126B cannot be easily determined through
this method as the trends break for objects later than M7. This revised distance is
somewhat larger than the distance estimated in \citet{Artigau2007} and
would revise upward the physical separation of the pair to about
6700~AU. Using the \citet{Reid1995} relation the TiO5 index alone points 
toward a slightly earlier spectral type (M$6.0\pm0.5$) for 2M0126A. 
This mild inconsistency in the distance to
2M0126AB illustrates the need for a more reliable distance measurement
based on parallax.

Using the GMOS $i$-band image of 2M0126AB, we rule out, using the
subtraction of a field stars' PSF, that either of its components is
itself an equal-luminosity binary with a separation
$>$0.2$\arcsec$. An object $\sim$4$\arcsec$ North-West of
2M0126A was examined as a plausible close-in companion, $2.0$~mag
fainter than 2M0126A. This object is undetected in $J$-band images
taken in 2006 June, implying a flux ratio greater than
$\sim$$3.0$~mag. This object is thus at least 1.0~mag bluer in
$i_{\rm{AB}}-J$ than 2M0126A and therefore cannot realistically be a
later-type companion. No other object, down to a $\Delta i_{\rm{AB}}
\sim$$4.5$~mag, is seen within $5\arcsec$ of either
component. The flux ratio between 2M0126A and 2M0126B is measured to
be $\Delta~i=0.518\pm0.015$~mag, consistent with, but much more
accurate, than the SuperCosmos Sky Survey \citep{Hambly2001} $I$-band photometry. A separation of
$81.89\pm0.2\arcsec$ is measured between the two components,
consistent with previous measurements.

Overall, aside from the fact that they compose the widest VLM
binary, 2M0126A and 2M0126B are basically indistinguishable from the
bulk of field M dwarfs, in age, surface gravity, activity level or
metallicity. The fact that 2M0126AB is a relatively old system makes
it all the more remarkable as compared to the bulk of wide VLM
binaries found in star-forming regions. This object demonstrates that
at least a handful of dynamically very fragile systems do survive to
become field objects; a result that future work on the dynamics of
star-forming regions will have to take into account.

The properties of 2M0126AB are interresting to compare to those of 
their more massive counterparts. The correlation between the upper limit on 
separation versus the mass of the primary (2500 $M_{\rm primary}^{1.54}$) noted by \citet{Abt1988} in
the B5-K0 interval clearly doesn't hold for the system described here, where we would expect a maximum 
separation of about 60~AU. Interrestingly, the order of magnitude of this limit is 
consistent with the sharp cutoff at $\sim50$~AU in the discribution of VLM binaries. While the 
discovery of 2M0126AB shows that much wider VLM binaries exist, the next step in the study 
of such systems would be a census of similar systems through a dedicated  survey, in which high care is taken 
in assessing the sensitivity and completeness limits such that clear and robust conclusions can be drawn. As these systems are likely to be 
very rare, a large sample of mid-to-late-Ms would be needed. It is also noteworthy that among the 500 or 
so L and T dwarfs in the 2MASS catalog, no binary wider than $2^{\prime\prime}$ has been identified. How rare are pairs with total masses
much below that of 2M0126AB at similar separations, if they exist at all, remains to be seen.

An interesting avenue of investigation will also be to determine whether forming pairs 
like 2M0126AB is fundamentally different from forming VLM binaries 
separated by tens of AUs. One striking property of VLM binaries, as compared to stellar binaries, 
is the strongly peaked mass ratio distribution toward unity \citep{Burgasser2007}, with 77\% of the systems 
having M$_2$/M$_1 \geq 0.8$, while late-F to K dwarfs binaries show a much flatter distribution. 
Interestingly, 2M0126AB has M$_2$/M$_1 \sim 0.95$, more in line with the VLM binary distribution. While 
obviously no broad-ranging conclusions can be drawn from the discovery of a single object, the discovery 
of only a handful of systems similar to 2M0126AB should clarify if they share the formation mechanisms of 
their siblings and represent the tail of the distribution, possibly dynamically excited during 
their formation, or if the formation scenarios themselves differ.

\acknowledgments
Based on observations obtained at the Gemini Observatory (program ID :
GS-2007B-Q-18), which is operated by the Association of Universities for Research in Astronomy, Inc., under a cooperative agreement
with the NSF on behalf of the Gemini partnership: the National Science Foundation (United
States), the Science and Technology Facilities Council (United Kingdom), the
National Research Council (Canada), CONICYT (Chile), the Australian Research Council
(Australia), Minist\'erio da Ci\^encia e Tecnologia (Brazil) and SECYT (Argentina).
DL is supported via a postdoctoral fellowship from the Fonds qu\'eb\'ecois de la recherche sur la nature 
et les technologies. RD is financially supported via a grant from the 
National Research and Engenerring Council of Canada. This publication has made use of the VLM Binaries Archive maintained 
by Nick Siegler at \hbox{http://www.vlmbinaries.org}.

\clearpage
\clearpage

\begin{center}

\begin{deluxetable}{lr@{$\pm$}lr@{$\pm$}l}
\tablewidth{0pt}
\tablecaption{Parameters of 2MASS J0126AB\label{tbl-1}}
\tablehead{
Parameter &\multicolumn{2}{c}{2M0126A}&\multicolumn{2}{c}{2M0126B}}
\startdata

2MASS designation&\multicolumn{2}{c}{J012655.49-502238.8}&\multicolumn{2}{c}{J012702.83-502321.1}\\
Angular separation$\,^{\rm a}$ [$\arcsec$]&\multicolumn{4}{c}{$81.86\pm0.10$}\\
$\mu_\alpha \cos\delta\,^{\rm a}$ [mas/yr]&$131$&$9$&$135$&$9$\\
$\mu_\delta\,^{\rm a}$ [mas/yr]&$-53$&$15$&$-47$&$15$\\
$V\,^{\rm a}$&\multicolumn{2}{c}{21.8}&\multicolumn{2}{c}{22.2}\\
$I\,^{\rm b}$&$17.2$&$0.3$&$17.6$&$0.3$\\
$\Delta i_{\rm{AB}}$&\multicolumn{4}{c}{$0.518\pm0.015$}\\
$J\,^{\rm c}$&$14.61$&$0.04$&$14.81$&$0.05$\\
$H\,^{\rm c}$&$14.05$&$0.05$&$14.16$&$0.04$\\
$K_s\,^{\rm c}$&$13.68$&$0.05$&$13.62$&$0.05$\\
FeH 1.20~$\mu$m$\,^{\rm a}$ [\AA]&$6.0$&$1.8$& $14.5$&$1.9$\\
K\textsc{I}~1.25$\mu$m$\,^{\rm a}$ [\AA]&$9.8$&$1.1$&$10.3$&$1.0$\\
H$\alpha$ pEW [\AA]&$-3.44$&$0.40$&$-7.32$&$0.50$\\
$\log(L_{\rm{H}\alpha}/L_{\rm{bol}})$&\multicolumn{2}{c}{$-4.5$}&\multicolumn{2}{c}{$-4.4$}\\
Li 671 nm [\AA]&\multicolumn{2}{c}{\textless$0.27$}&\multicolumn{2}{c}{\textless$0.52$}\\
Na I EW [\AA]&$7.04$&$0.12$&$6.69$&$0.20$\\
CaOH$\,^{\rm d}$&$0.32$&$0.05$&$ 0.46$&$0.11$\\
H$\alpha$$\,^{\rm d}$&$1.48$&$0.03$&$ 2.16$&$0.13$\\
CaH1$\,^{\rm d,g}$&$0.87$&$0.07$&$ 0.94$&$0.11$\\
CaH2$\,^{\rm d,g}$&$0.277$&$0.007$&$ 0.270$&$0.012$\\
CaH3$\,^{\rm d,g}$&$0.581$&$0.009$&$ 0.563$&$0.017$\\
TiO-a$\,^{\rm d,e}$&$2.71$&$0.09$&$ 2.81$&$0.16$\\
TiO2$\,^{\rm d}$&$0.32$&$0.02$&$ 0.33$&$0.03$\\
TiO3$\,^{\rm d}$&$0.55$&$0.03$&$ 0.54$&$0.04$\\
TiO4$\,^{\rm d}$&$0.53$&$0.03$&$ 0.71$&$0.06$\\
TiO5$\,^{\rm d,g}$&$0.219$&$0.008$&$ 0.284$&$0.016$\\
VO-a$\,^{\rm d,e}$&$1.073$&$0.006$&$ 1.104$&$0.011$\\
VO-b$\,^{\rm e}$&$1.256$&$0.009$&$ 1.361$&$0.013$\\
PC3$\,^{\rm d}$&$1.569$&$0.008$&$ 1.850$&$0.014$\\
Near-IR$\,^{\rm a}$ \& optical SpT&M$6.5$V&$0.5$&M$8$V&$0.5$\\
Photometric distance$\,^{\rm a}$ [pc]&63&$5$ pc&61&$6$\\
Physical separation$\,^{\rm a}$ [AU]&\multicolumn{4}{c}{5100$\pm$400}\\
$T_{eff}\,^{\rm a}$ [K]&$2670$&$180$&$2490$&$180$\\
Mass ($200$~Myr)$\,^{\rm h}$ [M$_{\odot}$]&$0.068$&$0.005$&$0.065$&$0.005$\\
Mass ($>1$~Gyr)$\,^{\rm h}$ [M$_{\odot}$]&$0.095$&$0.005$&$0.092$&$0.005$\\
Age [Myr]&\multicolumn{4}{c}{$\textgreater$200}\\
M$_{\textrm{bol}}\,^{\rm a}$&$12.76$&$0.14$&$12.75$&$0.14$\\
Luminosity$\,^{\rm a}$ [$\log \left( \textrm{L}/\textrm{L}_\odot \right)$]&$-3.21$&$0.05$&$-3.20$&$0.05$\\
\enddata
\tablenotetext{a}{From \citet{Artigau2007}, see this reference for more details.}
\tablenotetext{b}{From the SuperCosmos Sky Survey catalogue.}
\tablenotetext{c}{From the 2MASS point source catalogue.}
\tablenotetext{d}{\citet{Cruz2002} index}
\tablenotetext{e}{\citet{Kirkpatrick1999} index}
\tablenotetext{f}{\citet{Martin1999} index}
\tablenotetext{g}{\citet{Gizis1997} index}
\tablenotetext{h}{Determined using the \citet{Chabrier2000} evolution models and the estimated $J$, $H$ and $K_s$ absolute magnitudes.}
\end{deluxetable}

\end{center}

\end{document}